\documentclass[a4paper]{jpconf}
\usepackage{graphicx}
\usepackage{xspace}

\begin{document}
\title{APEIRON: composing smart TDAQ systems for high energy physics experiments}

\author {
Roberto~Ammendola$^2$,
Andrea~Biagioni$^1$,
Carlotta~Chiarini$^{3,1}$,
Andrea~Ciardiello$^{3,1}$,
Paolo~Cretaro$^1$,
Ottorino~Frezza$^1$,
Francesca~Lo~Cicero$^1$,
Alessandro~Lonardo$^1$,
Michele~Martinelli$^1$,
Pier~Stanislao~Paolucci$^1$,
Pierpaolo Perticaroli$^1$,
Cristian~Rossi$^1$,
Francesco~Simula$^1$,
Matteo~Turisini$^1$,
Piero~Vicini$^1$
}

\address{$^1$ Istituto Nazionale di Fisica Nucleare (INFN), sezione di Roma, Rome, Italy}
\address {$^2$ Istituto Nazionale di Fisica Nucleare (INFN), sezione di Roma Tor Vergata, Rome, Italy}
\address {$^3$ Dipartimento di Fisica, Sapienza Universit\`a di Roma, Rome, Italy}

\ead{alessandro.lonardo@roma1.infn.it}

\begin{abstract}


We present APEIRON, a distributed heterogeneous processing framework comprising both hardware architecture and software stack for multi-FPGA systems. Targeting smart trigger and data acquisition (TDAQ) systems in high energy physics, APEIRON spans the full software hierarchy: from low-level device drivers to a high-level dataflow programming model based on High-Level Synthesis. We describe the framework design, its core communication infrastructure, and a particle identification application for the NA62 experiment as a representative physics use case.
\end{abstract}

\section{Motivation and design goals}
Real-time dataflow processing in experimental particle physics places stringent demands on computing throughput, deterministic latency and I/O bandwidth. FPGA devices are particularly well suited to these requirements owing to their reconfigurable logic, tightly coupled memory and high-speed serial transceivers. The maturation of High-Level Synthesis (HLS) tools over the past decade has substantially lowered the entry barrier, allowing a wider community of physicists and engineers to exploit FPGA acceleration without resorting exclusively to Hardware Description Language workflows.

A significant limitation of present-day HLS environments, however, is their confinement to a single FPGA device. When the scale of a trigger or data-acquisition problem exceeds the capacity of one chip—as is common in modern experiments with high channel counts and event rates—developers must fall back on ad-hoc, manually crafted inter-FPGA communication layers. This gap motivated the creation of APEIRON: an integrated framework that extends the Xilinx Vitis HLS ecosystem to operate transparently across a network of interconnected FPGAs.

The principal design goals can be summarised as follows:
\begin{itemize}
    \item provide a modular, topology-configurable, low-latency direct interconnect among FPGA nodes;
    \item offer a dataflow programming abstraction, drawing on Kahn Process Networks~\cite{gilles1974semantics}, in which processing tasks communicate through lightweight send/receive primitives regardless of their physical placement;
    \item automate the generation of all ancillary logic (routing, dispatching, aggregation) from a compact application description, so that users concentrate on the algorithmic C/C++ kernels;
    \item{target both traditional low-level trigger systems and data-reduction stages in trigger-less or streaming readout experimental setups characterised by high event rates.}
\end{itemize}

\section{Platform architecture}

\subsection{Overall topology}
\begin{figure}[hbt!]
\begin{center}
\includegraphics[width=0.6\textwidth]{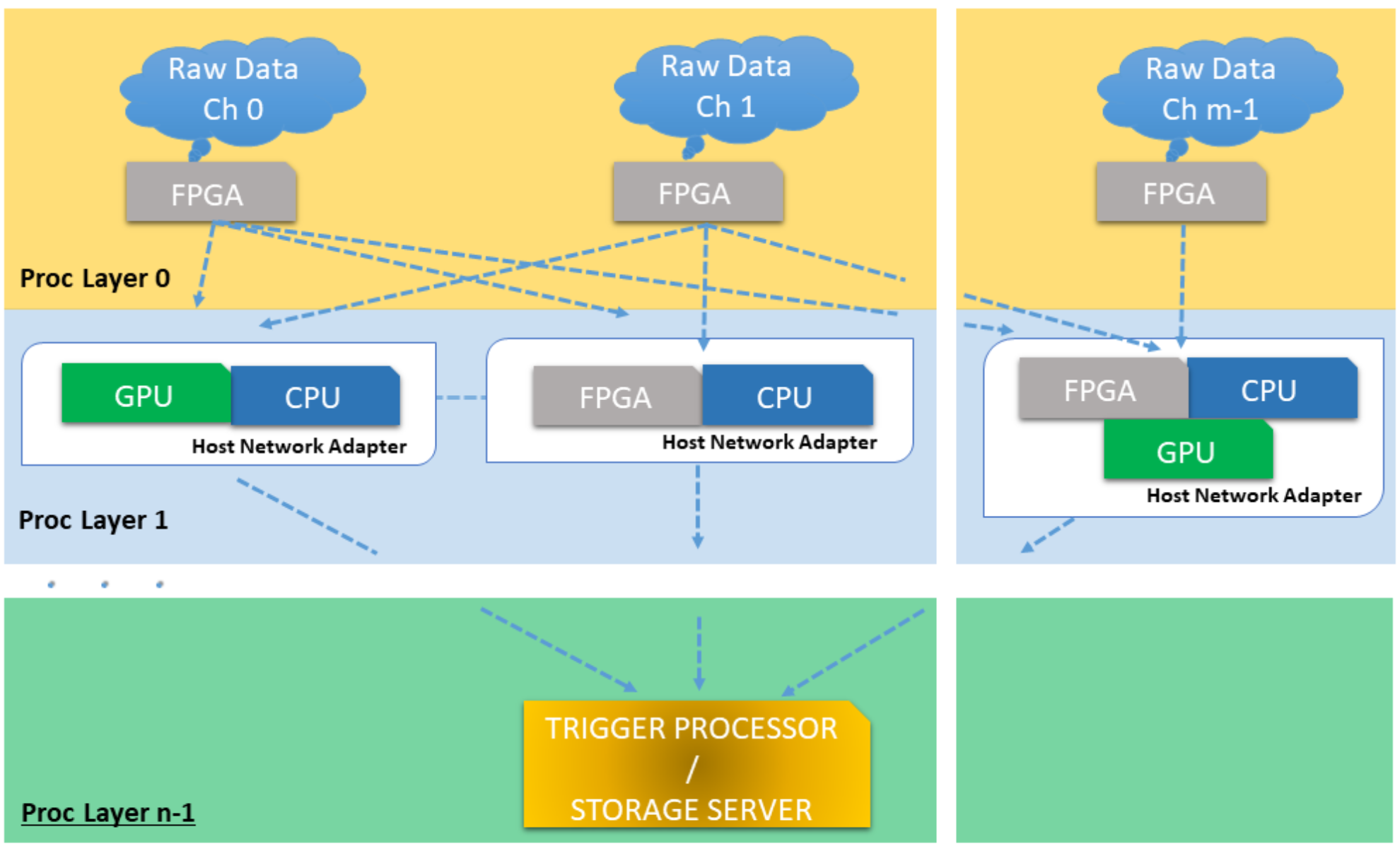}
\end{center}
\caption{\label{proc_layers}Data-stream recombination across specialised processing stages in a trigger or data-reduction pipeline built with APEIRON.}
\end{figure}

At the system level, APEIRON models the data path of a trigger or data-reduction chain as \textit{m} independent data sources (detectors or sub-detectors) feeding a cascade of \textit{n} stream-processing stages. Each stage may reside on one or more FPGA nodes; the network fabric recombines data streams across stages as required by the physics algorithm. Figure~\ref{proc_layers} illustrates a representative configuration.

The scalability of this scheme rests on the network infrastructure: a packet-switched, dimension-order-routed mesh whose physical topology can be tailored to the application through configuration rather than redesign. The network constitutes the core enabling element of the platform, bridging the gap between single-device HLS and multi-device real-time systems.

\subsection{Communication infrastructure}
The communication subsystem descends from the APEnet~\cite{APEnetTwepp:2013} and EXANEST~\cite{DSD:EXANEST:2017} network-on-chip designs originally developed for HPC, adapted here for low-latency stream processing in TDAQ applications. The Communication IP was co-designed with the APEIRON software stack to achieve low latency and scalable bandwidth between processing tasks.

\begin{figure}[hbt!]
\begin{center}
\includegraphics[width=0.7\textwidth]{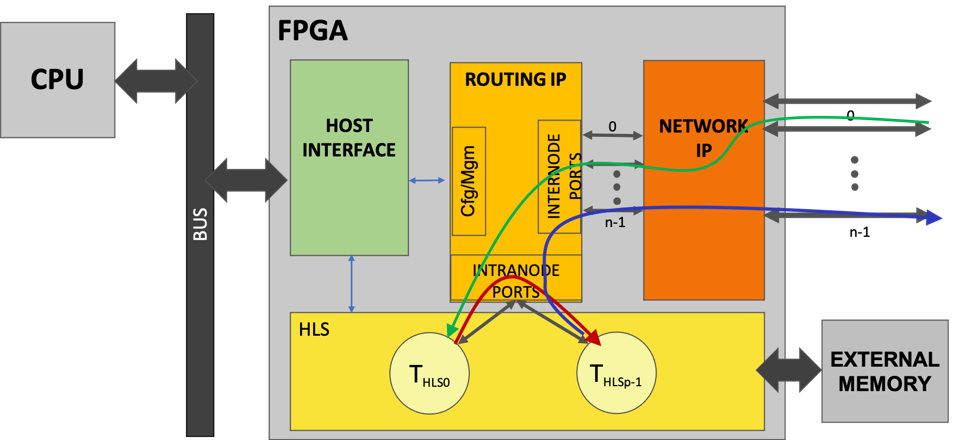}
\end{center}
\caption{\label{node}Schematic of intra-node (red) and inter-node (green: receive; blue: send) data paths between HLS kernels.}
\end{figure}

Two classes of data movement are supported (see Figure~\ref{node}):
\begin{description}
    \item[Intra-node] transfers between kernels co-located on the same FPGA, routed through the local switch fabric.
    \item[Inter-node] transfers between kernels residing on different FPGAs, carried over high-speed serial transceivers and routed by the network.
\end{description}

\noindent\textbf{Routing and switching.}
The Routing IP contains a crossbar switch, configuration/status registers and the inter-node and intra-node interfaces. Paths across the switch are set up dynamically by a router that applies dimension-order routing—offsets are resolved one coordinate at a time—while an arbiter resolves contention when multiple packets compete for the same egress port. The switching technique is Virtual Cut-Through~\cite{Kermani79virtualcut-through:}: forwarding of a packet begins as soon as the output direction is determined and sufficient buffer space is available. Deadlock freedom under dimension-order routing is ensured by providing two virtual channels per physical link~\cite{Duato:1995:Deadlock}; fault tolerance is not currently addressed.

All transfers are packet-based. Each packet comprises a header carrying routing and addressing metadata, a variable-length payload and a trailing footer.

\subsection{Kernel interface and programming model}
To keep the developer's task focused on the algorithm, APEIRON imposes only a minimal requirement on HLS kernels: each must expose
AXI4-Stream~\cite{AmbaAxiStreamProtSpec:2021} ports for its input and output channels, following the prototype:

\begin{verbatim}
void my_task(
    [optional kernel-specific parameters],
    message_stream_t message_data_in[N_INPUT_CHANNELS],
    message_stream_t message_data_out[N_OUTPUT_CHANNELS]);
\end{verbatim}

Communication between any two kernels—co-located or remote—is expressed through a lightweight C++ API consisting of a non-blocking \texttt{send()} and a blocking \texttt{receive()}:

\begin{verbatim}
size_t send(msg, size, dest_node, task_id, ch_id);
size_t receive(ch_id);
\end{verbatim}

\noindent Here \texttt{dest\_node} encodes the multi-dimensional coordinate of the target FPGA, \texttt{task\_id} selects the receiving kernel within that node (up to four per node), and \texttt{ch\_id} identifies one of up to 128 logical channels local to the receiving kernel. The API exploits AXI4-Stream side-channel signals to convey all the information needed to construct the packet header, allowing the developer to perform inter-kernel communication without knowledge of the underlying packet protocol.

\begin{figure}[hbt!]
\begin{center}
\includegraphics[width=0.7\textwidth]{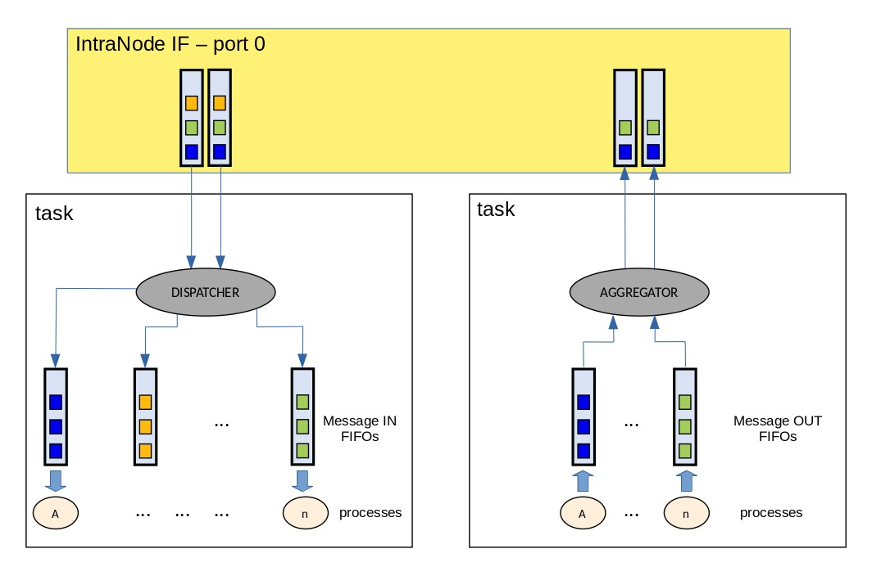}
\end{center}
\caption{\label{disp_aggr}Dispatcher/Aggregator pair interfacing Intra-node Port~0 to the associated HLS task. Individual message-input FIFOs are selected by the \texttt{ch\_id} API parameter.}
\end{figure}

\begin{figure}[hbt!]
\centering
  \begin{minipage}[t]{.65\textwidth}
\centering
    \includegraphics[width=.95\textwidth]{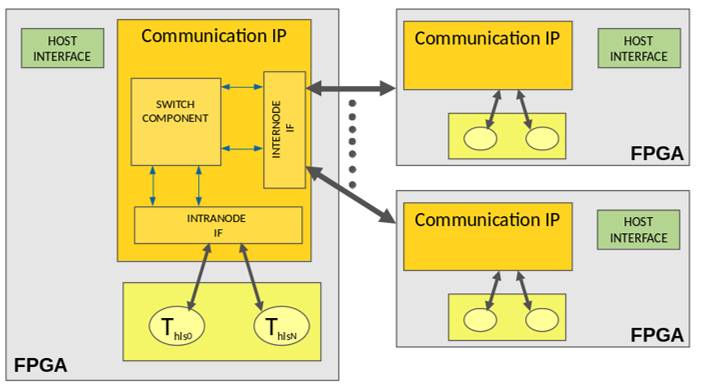}
    \caption{Communication IP connecting HLS kernels (yellow ovals) through data-stream I/O and inter-kernel messaging.}
    \label{fig:comm_ip}
  \end{minipage}
  \quad
  \begin{minipage}[t]{.25\textwidth}
\centering
    \includegraphics[width=.95\textwidth]{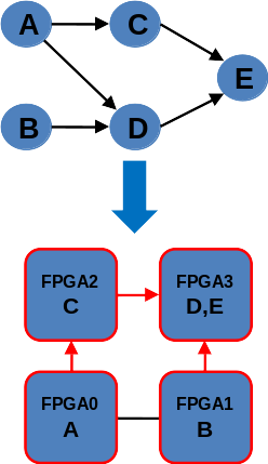}
    \caption{Example mapping of five processing tasks onto four networked FPGAs.}
    \label{fig:mapping}
  \end{minipage}
\end{figure}

Within each node, the Communication IP connects the user kernels to the Routing IP through two dedicated adapter blocks (Figure~\ref{disp_aggr} and Figure~\ref{fig:comm_ip}):
\begin{itemize}
    \item the \textbf{Dispatcher}, which inspects the header of incoming packets and steers them to the appropriate input channel FIFO of the target kernel;
    \item the \textbf{Aggregator}, which accepts outgoing data from the kernel, constructs the packet header and enqueues the result into the Routing IP's transmission buffers.
\end{itemize}

\subsection{Build flow}
The mapping of the computational dataflow graph onto the physical FPGA network (illustrated in Figure~\ref{fig:mapping}) is specified in a YAML configuration file that lists, for every kernel, its input/output channel count and the intra-node port to which it is attached. From this description the APEIRON toolchain links the Communication and Routing IPs, wires them to the user kernels and produces the complete design bitstream. The designer is thus relieved of manual integration work and can iterate on the algorithmic content of the C/C++ kernels alone.

\section{Validation: particle identification for NA62}

\subsection{Experimental context}
NA62 is a fixed-target kaon-decay experiment operating at the CERN SPS North Area \cite{Gil_2017NA62DetectorAndBeamline}.
Its Ring Imaging Cherenkov (RICH) detector provides information that can be used for online particle identification (PID), which is a valuable input to the low-level trigger decision.

An earlier incarnation of the system, dubbed GPURICH, offloaded a geometry-based PID algorithm to a GPU, with a dedicated FPGA board running the NaNet~\cite{NanetTwepp:2013,NaNet:TWEPP2014:paper,NanetTwepp:2015} firmware to achieve direct, low-latency data transfer between the detector readout and GPU memory. In the present work we replace this heterogeneous GPU+FPGA pipeline with a single-FPGA solution—termed FPGA-RICH—built entirely within the APEIRON framework, using HLS.

\subsection{Neural-network model}
FPGA-RICH ingests RICH events in a continuous stream and must sustain a throughput exceeding 10\,MHz, as dictated by experiment requirements. The PID task is cast as a neural-network inference problem using a \emph{seedless} approach: the network receives only RICH hit data, without external track seeds.

The chosen architecture (Figure~\ref{DenseModel}) is a fully connected network with three layers of 64, 16 and 4 neurons respectively. Up to 64 normalised photomultiplier identifiers per event serve as input features. To constrain FPGA resource usage, the trained model undergoes quantisation with QKeras~\cite{Coelho:2724942} and is subsequently converted to synthesisable firmware using HLS4ML~\cite{Duarte_2018}, yielding fixed-point representations of $<$8,~1$>$ for weights and biases
and $<$16,~6$>$ for activations.

Two quantities are inferred per event: the multiplicity of charged particles ($N_r$) and the count of $e^{\pm}$ ($N_e$). Training and validation datasets were assembled from NA62 physics runs using the experiment's offline analysis framework; ground-truth labels were obtained from the seedless \emph{RichReco} reconstruction algorithm.

\subsection{Training strategy and results}
\begin{figure}[hbt!]
\centering
  \begin{minipage}[t]{.45\textwidth}
\centering
    \includegraphics[width=1.3\textwidth]{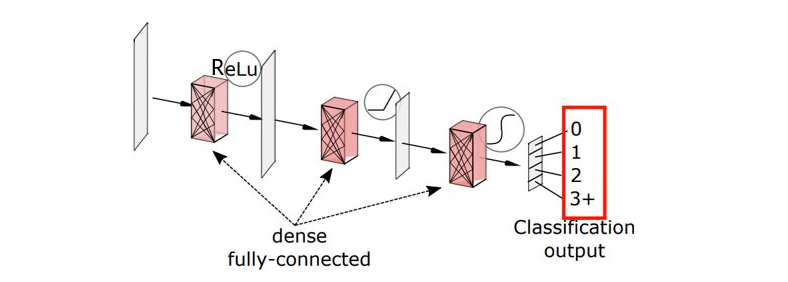}
    \caption{Topology of the dense neural-network model (64\,$\times$\,16\,$\times$\,4).}
    \label{DenseModel}
  \end{minipage}
  \quad
  \begin{minipage}[t]{.5\textwidth}
\centering
    \includegraphics[width=0.9\textwidth]{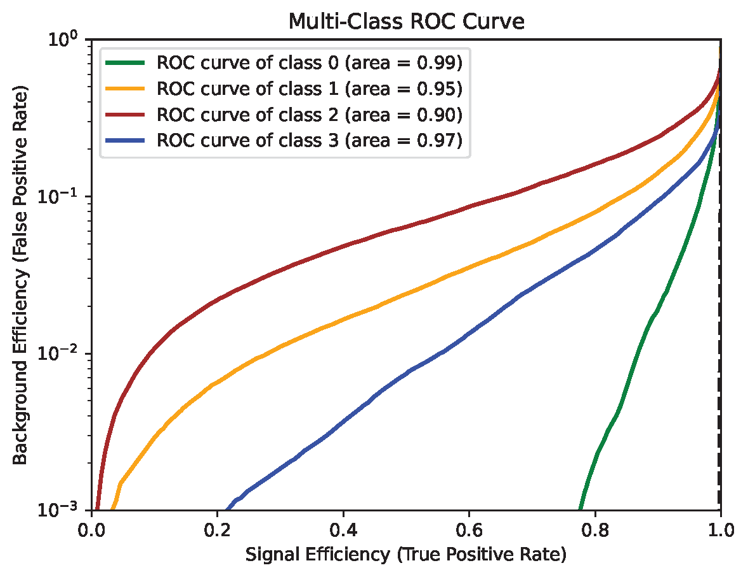}
    \caption{Receiver operating characteristic curves for $N_r$ classification.}
    \label{ROC}
  \end{minipage}
\end{figure}

Because the network output is intended to drive a trigger decision, inference performance is critical. The training set comprised approximately 3 million events from run~8011, chosen to be representative of online conditions. An independent validation set of 3.5 million events from run~8893 was used to assess generalisation. The ROC curves for $N_r$ classification, shown in Figure~\ref{ROC}, demonstrate satisfactory discriminating power.

Since the NA62 RICH detector is able to discriminate the type of charged particles only in the $15-35$~$GeV/c$ energy range, results for $N_e$ are not equally satisfying.

\subsection{Resource utilisation and throughput}
Synthesis targeting a Xilinx VCU118 board at 150\,MHz yielded a very compact implementation consuming 14\,\% of available LUTs and 2\,\% of DSP slices. The measured inference latency is 146.66\,ns, corresponding to a sustained throughput of 18.75\,MHz—comfortably above the 10\,MHz requirement.

\section{Conclusions and Future Work}
We have presented the APEIRON framework, whose combination of a configurable low-latency FPGA network, a dataflow programming model and automated build tooling addresses a recurrent need in the design of trigger and data-acquisition systems for particle physics.

We are continuing the development of the APEIRON framework in order improve its performance and usability. We are also finalising the development of the FPGA-RICH system within the framework, encouraged by the good performance on the identification of charged particles. We envision a solution to improve results in identification of $e^{\pm}$, using the LKr calorimeter online primitives that provide information related to the energy of the event.

\section*{References}
\bibliography{bibliography}

\end{document}